\documentclass{optica-article}

\journal{opticajournal} 

\articletype{Research Article}

\usepackage{graphicx}
\usepackage{dcolumn}
\usepackage{bm}
\usepackage{physics}
\usepackage{dsfont}
\usepackage{cite} 
\usepackage{lineno}
\usepackage{float} 

\begin{document}

\title{Experimental demonstration of a multi-particle collective measurement for optimal quantum state estimation}

\author{Arman Mansouri,\authormark{1,*} Kyle M. Jordan,\authormark{1} Raphael A. Abrahao,\authormark{1,2} Jeff S. Lundeen\authormark{1}}

\address{\authormark{1}Department of Physics and Max Planck Centre for Extreme and Quantum Photonics, University of Ottawa, 25 Templeton Street, Ottawa, Ontario, Canada K1N 6N5\\
\authormark{2}School of Physics and Astronomy, Future Photon Initiative, Center for Detectors, Rochester Institute of Technology, Rochester, NY 14623, USA }

\email{\authormark{*}amans069@uottawa.ca} 


\begin{abstract*} 
We experimentally demonstrate a two-particle collective measurement proposed as the optimal solution to a quantum state estimation game. Our results suggest that, in practice, the collective measurement strategy is at least as good as the best local approach, and it achieves a higher average fidelity when accounting for systematic errors. This photonic implementation uses a recently developed universal two-photon projective measurement based on Hong-Ou-Mandel interference, polarization-dependent loss, and unitary operations. We compare the performance to the case where the entangling component of the measurement is suppressed. We further apply the collective measurement to quantum state tomography, observing a near-optimal scaling of the infidelity with the total number of samples. 
\end{abstract*}

\section{Introduction}
Collective measurements probe many-body properties such as total spin or permutation symmetry rather than properties such as single-particle spin. They arise naturally in many-body systems, for instance in the study of ferromagnetism, where the macroscopic field is a collective effect of the many spins. Collective measurements are nonlocal in the sense that they project onto entangled states and, thus, they enable applications that are impossible within the paradigm of Local Operations and Classical Communication (LOCC). Probing collective properties provides optimal solutions to several parameter and state estimation problems in quantum information \cite{Holevo, MassarPopescu1995, derka_universal_1998, latorre_minimal_1998, vidal1999, brus_optimal_1999, Nagaoka, hayashi_reexamination_2005, Liang2006, Bagan2006, Tsang2011, Munoz2018, Wu2020, conlon2023, conlon_discriminating_2023 ,Crossman_2024, Conlon2025}. However, the majority of implementations in state estimation and metrology have so far been focused on the use of local measurements, due to the practical challenges associated with the realization of collective measurements. Some photonic implementations have relied on encoding the qubits in different degrees of freedom of the same photon \cite{Hou2018, zhou_experimental_2025}. While this seemingly simplifies the task of performing collective measurements, the outcomes can be explained by classical optics. In particular, outcome statistics can be predicted based on first-order interference. Moreover, the states corresponding to the measurement projector do not allow useful Bell violations and quantum information processing tasks. Viable experimental methods for collective measurements will open up the above-mentioned optimal sensing and state estimation applications. 

An important and fundamental task in quantum information is quantum state tomography, which allows complete characterization of the state $\rho$ of quantum systems through measurements on an identically prepared ensemble of size $N_{\text{ens}}$. This procedure involves the measurement of a set of observables sufficient for the complete determination of the density matrix, resulting in a state estimate $\rho_\text{est}$ \cite{DAriano}. For a $d$-dimensional system, at least $d^2 - 1$ projection measurements are needed for a tomographic reconstruction. In practice, many samples need to be measured for each projection to reduce statistical uncertainty, meaning a large total number of samples are required for quantum state tomography. This poses a challenge for the efficient characterization of quantum systems. A common way to assess the efficiency of a state characterization scheme is to consider the scaling of the infidelity
    $1 - F = 1 - \left[\text{Tr} \sqrt{\sqrt{\rho_\text{est}}\rho \sqrt{\rho_\text{est}}}\right]^2$
with the number of measured samples $N_\text{ens}$, which quantifies the error in the estimate $\rho_\text{est}$ of the true state $\rho$. Tomographic scaling is important from a fundamental point of view, since it relates to how much information can be extracted. It is also significant when considering the practicality in experiments where the number of copies is limited.

Through a calculation based on the quantum Fisher information matrix and the quantum Cramér-Rao bound, Gill and Massar showed that the optimal infidelity for the tomography of a pure one-qubit state scales as $1/N_\text{ens}$ \cite{Gill-Massar}. Massar and Popescu discussed a related state estimation task, where $N$ qubits polarized in $\ket{\vec{n}}$ are provided in each round, and a single guess $\ket{\vec{n}_g}$ of the polarization vector is made \cite{MassarPopescu1995}. 

We describe two experiments on state identification, starting with the game addressed by Massar and Popescu, and later applying the collective measurement to quantum state tomography. The Massar and Popescu game concerns state estimation in a scenario where the size of the ensemble may not be large (statistical). The state is thus identified through a guess corresponding to each measured outcome. We examine two variants of the game: one with uniformly distributed unknown states, and the other with the unknown state being one of four equiprobable states. When the number of available copies is sufficiently large, tomographic estimation becomes possible, with a continuous range of estimated states rather than discrete guesses. The optimal two-copy collective measurement of the Massar and Popescu game is then used to tomographically estimate the state from a large ensemble in pairs, allowing to examine its performance in the statistical regime.

\subsection{Optimal Massar and Popescu game} The optimal measurement for the Massar and Popescu game is defined as one that, on average, maximizes the fidelity $\mathcal{F}(\ket{\vec{n}}, \ket{\vec{n}_g}) = \abs{\braket{\vec{n}}{\vec{n}_g}}^2$. The procedure is repeated many times, with $\ket{\vec{n}}$ given according to some prior distribution $P(\vec{n})$, e.g., uniformly distributed over the Bloch sphere. The average fidelity $\bar{\mathcal{F}}$ can thus be written as \cite{MassarPopescu1995, GisinPopescu1999}:
\begin{equation}
\begin{aligned}
		\bar{\mathcal{F}} &= \int   P(\vec{n}) \ d\vec{n} \sum_g P(g|\vec{n}) \mathcal{F}(\ket{\vec{n}}, \ket{\vec{n}_g}) \\
  &= \int  P(\vec{n}) \ d \vec{n} \sum_g P(g|\vec{n})\left(\frac{1+ \vec{n}\cdot    \vec{n}_g}{2}\right), \label{avgFidExpression}
  \end{aligned}
\end{equation}
where $ \vec{n} $ is the Bloch vector of the unknown state $ \ket{\vec{n}} $, $ \vec{n}_g $ is the Bloch vector associated with the guessed state, and $ P(g|\vec{n}) $ is the probability of guessing $ \vec{n}_g $ when the true direction is $ \vec{n} $.

Massar and Popescu found that when the different directions $\vec{n}$ are uniformly distributed over the unit sphere, i.e., $P(n)$ is uniform, the optimal measuring strategy is to perform a collective measurement on the ensemble of $N$ qubits. This strategy maximizes the average fidelity to $\bar{\mathcal{F}}^{\text{Coll}}_\text{GenMP} = (N+1)/(N+2)$ \cite{MassarPopescu1995}. For the case of $N=2$, an optimal measuring strategy is to measure a nondegenerate operator whose eigenstates are the following superpositions of the singlet state $ \ket{S}= \frac{1}{\sqrt{2}}(\ket{01} - \ket{10})$ and product tetrahedron states $\ket{\uparrow_{n_i}\uparrow_{n_i}}$:
\begin{equation}
    \begin{aligned}
    \ket{\text{MP}_1} &= \frac{1}{2}\ket{S} + \frac{\sqrt{3}}{2} \ket{\uparrow_{n_1}\uparrow_{n_1}}\\ \ket{\text{MP}_j} &= -\frac{1}{2}\ket{S} + \frac{\sqrt{3}}{2} \ket{\uparrow_{n_j}\uparrow_{n_j}}, \quad j=2,3,4. \label{e-states}
    \end{aligned}
\end{equation}
In the Bloch sphere, the states $\ket{\uparrow_{n_i}}$ are located on the surface of the sphere at the vertices of a tetrahedron and can be chosen as:
\begin{equation}
    \begin{aligned}
    \ket*{\uparrow_{n_1}} &= \ket{0}; \\
    \ket*{\uparrow_{n_j}} &= \frac{1}{\sqrt{3}}\ket{0} + \sqrt{\frac{2}{3}}e^{i\alpha_j}\ket{1}, \qquad j=2,3,4, \label{eq:tetras}
    \end{aligned}
\end{equation}
with $\alpha_2,\alpha_3,\alpha_4 = 0, \ 2\pi/3, \ 4\pi/3$. Upon performing this collective measurement and obtaining the outcome $i$ (corresponding to $\ket{\text{MP}_i}$), the tetrahedron state $\ket{\uparrow_{n_i}}$ is chosen as the guess $\ket{\vec{n}_g}$. Projections onto $\ket{\text{MP}_i}$ require collective measurements, since $\ket{\text{MP}_i}$ is entangled, albeit non-maximally.

The theoretical average fidelity of the Massar and Popescu game in different scenarios can be directly calculated from Eq. \ref{avgFidExpression}. If the unknown states are distributed along uniform directions on the unit sphere, $P(\vec{n}) d\vec{n} = \sin\theta d\theta d\phi/4\pi$ and the average fidelity using the optimal collective approach is $\bar{\mathcal{F}}^\text{Coll}_{N=2, \text{ GenMP}} = 3/4 = 0.75$. We will refer to this case as the collective approach to the General Massar and Popescu (GenMP) game. The average fidelity using the best local approach is $\mathcal{F}^\text{LOCC}_{{N=2}, \text{ GenMP}} = \frac{1}{6}(3+\sqrt{2}) \approx 0.7357$ \cite{Massar2000}. 

When the \textit{a priori} distribution $P(n)d\vec{n}$ is made discrete in such a way that $\vec{n}$ is restricted to being one of the tetrahedron states $\ket{\uparrow_{n_i}}$, we have what we refer to as the Tetrahedron Massar and Popescu (TetraMP) game. The collective approach to the TetraMP game gives an average fidelity of $\bar{\mathcal{F}}^\text{Coll}_{N=2,\text{ TetraMP}} = 5/6 \approx 0.8333$, conjectured to be optimal \cite{GisinPopescu1999}. To the best of the authors' knowledge, the optimal LOCC approach is not known for the TetraMP game. Alternatively, the TetraMP game can be treated as a quantum state discrimination problem for the equiprobable set of states $\{\ket{\uparrow_{n_i}}\}$. While lower and upper bounds for multiple-state, minimum-error discrimination using single-copy measurements have been derived \cite{loubenets_general_2022}, they may not be achievable or tight (yielding a wide average fidelity range $[ 0.636, 0.908]$ in this case). More concretely, the collective strategy can be compared to the ``pretty good" measurement \cite{belavkin1975optimal,hausladen1994pretty}, expected to perform fairly well for the discrimination of any set of states. The pretty good measurement corresponds to the POVM $\{\frac{1}{2}\ketbra{\uparrow_{n_i}}\}$ for the TetraMP game and yields an average fidelity of $\bar{\mathcal{F}}^\text{PGM}=2/3 \approx 0.6667$, which is $\sim 16.7\%$ lower than $\bar{\mathcal{F}}^\text{Coll}_{N=2,\text{ TetraMP}}$. 

The theoretical fidelity bounds outlined above will serve as benchmarks for evaluating our experimental results.

\section{Methods}
\subsection{Implementation of the Massar and Popescu measurement} 
Attempts to implement the Massar and Popescu game have so far been limited to the use of photonic quantum walks, where various degrees of freedom of the same photon encode different qubits \cite{Hou2018, zhou_experimental_2025}. Although this simulates the collective strategy using non-identical degrees of freedom of a single photon, such projection states cannot meaningfully violate Bell's inequalities or be applied to useful quantum computing. The same outcomes could be obtained with classical light of the same mean photon number, and the outcome statistics could be predicted without a quantum-mechanical treatment, based on first-order interference. Nonetheless, photonic quantum walks are valuable for simulating topological physics\cite{kitagawa2012observation}, quantum algorithms\cite{qiang2021implementing}, as well as providing a testing ground for collective measurements.

\begin{figure}[H]
    \centering
    \includegraphics[scale=0.43]{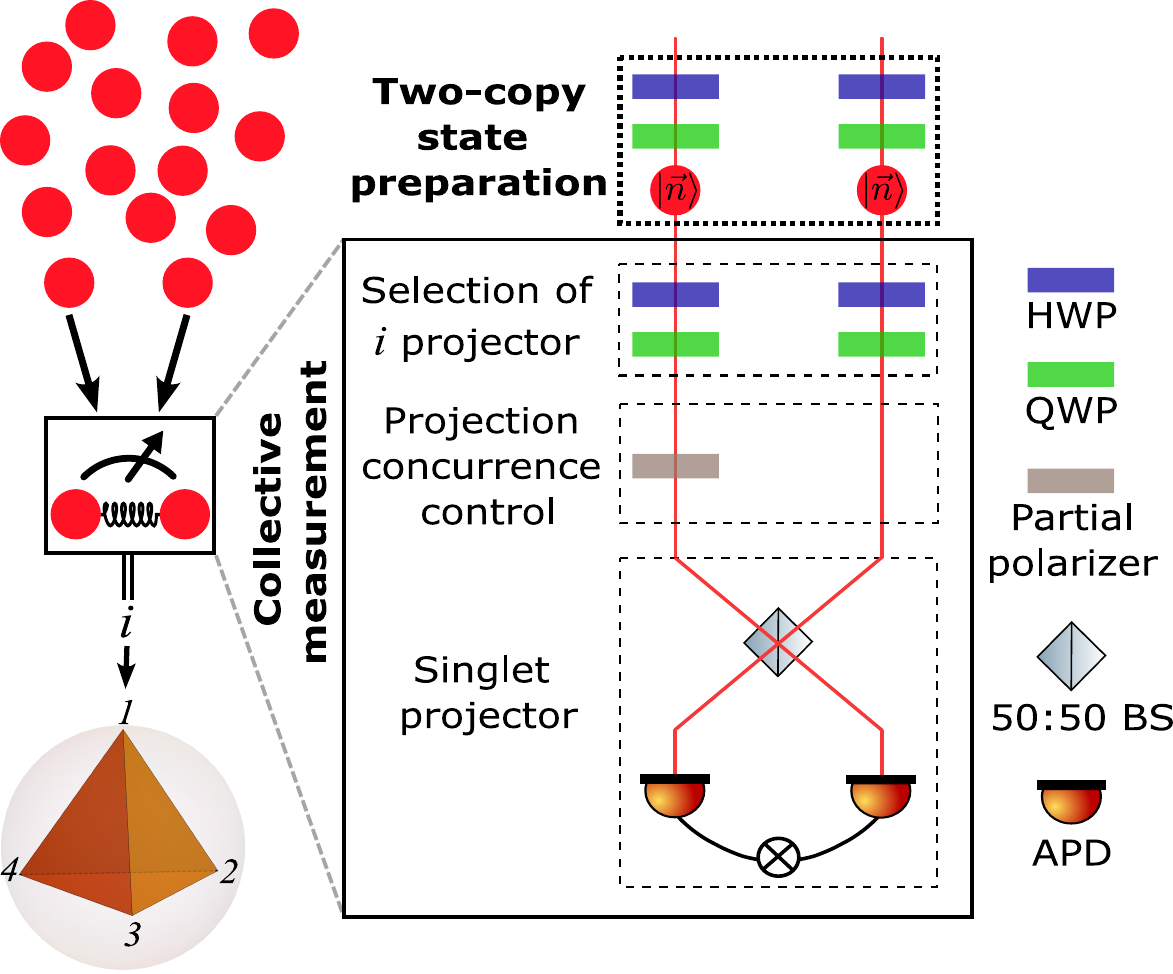}
    \caption{Scheme and experimental setup used for the Massar and Popescu game. The setup consists of half-wave plates (HWP), quarter-wave plates (QWP), a partial polarizer, a 50:50 beamsplitter (BS), and  single-photon avalanche photodiode detectors (APD). See the main text for details.}
    \label{fig:setup}
\end{figure}

Instead, we begin with two identical photons, which are our two copies in the ensemble for the Massar and Popescu game. Our multi-particle implementation involves an entangling measurement of two-photon polarization states that combines Hong-Ou-Mandel (HOM) interference~\cite{HOM_PRL87, PRA2022_Quantum-metrology-HOM} with polarization-dependent loss \cite{berry_fair-sampling_2010, ringbauer_information_2014}, as shown in Fig. \ref{fig:setup} \cite{Thekkadath_2018}.

We used a narrow-linewidth 40 mW 405 nm continuous wave laser to pump a 2 mm Type II BBO crystal. The orthogonally polarized 810 nm photon pairs were generated through collinear spontaneous parametric down-conversion and then were split at a polarizing beamsplitter and then coupled into polarization-maintaining fibers. The photons were out-coupled into two arms, with one out-coupler on a translation stage to vary the relative delay between the photons. In each arm a half-wave plate (HWP) and a quarter-wave plate (QWP) were used to prepare the input polarization state $\ket{\vec{n}}\otimes\ket{\vec{n}}$, with the encoding $\ket{0} \equiv \ket{H}$ and $\ket{1} \equiv \ket{V}$.

 Whereas typical two-photon polarization measurements project onto either a separable or a maximally entangled state, we use a recently developed general two-photon polarization measurement \cite{Thekkadath_2018} to project on the non-maximally entangled $\ket{\text{MP}_i}$ states. At the heart of our device is a projection onto the singlet state $\ketbra{S}{S}$: this is accomplished when two photons entering opposite input ports of a 50:50 beamsplitter exit opposite ports \cite{Bouwmeester1997}.  The device varies the level of entanglement (i.e, concurrence) of the projection state by introducing a partial polarizer in one input mode, with transformation matrix 
\begin{equation}
		\hat{W} = \begin{bmatrix}
			t_H & 0 \\ 0 & t_V
		\end{bmatrix} \otimes \mathds{I}_{2\times2},
	\end{equation}
where $t_H$ and $t_V$ represent complex transmission amplitudes, here real-valued. With the partial polarizer, we alter the projection onto $\ket{S}$ to have a lower entanglement, a concurrence $\mathcal{C}(\ket{\text{MP}_i}) = (2\abs{t_H}\abs{t_V})/(t^2_H+t^2_V)=0.25$ for all four $\ket{\text{MP}_i}$. In turn, this determines the efficiency of the projection $\eta = 1/(1+\sqrt{1-\mathcal{C}^2}) $, which is the probability of a successful projection given the incoming pair of photons were in the state $\ket{\text{MP}_i}$.
To project onto a specific $\ket{\text{MP}_i}$, we add an appropriate unitary transformation in each input mode. Altogether, this device projects onto any one of the states $\ket{\text{MP}_i}$. 

Two-photon interference has been used to demonstrate collective measurements for parameter estimation and metrology \cite{Roccia_2018, Parniak}. However, until the present work only basic two-photon collective measurements have been used in applications: projections on maximally entangled states. These projections are naturally implemented by a standard 50:50 beamsplitter. In this work, we project onto optimized collective states $|\text{MP}_i\rangle$, which are non-maximally entangled and non-trivial to implement.

 The two-photon projections are implemented experimentally as shown in Fig. \ref{fig:setup}. After the input states are prepared, a HWP and QWP implement the appropriate unitary on each input photon. The concurrence $\mathcal{C} =0.25$ corresponds to a polarization-dependent loss with extinction ratio $\sim 62:1$ between the vertical and horizontal polarization optical powers. Since this extinction ratio is significantly biased towards one polarization, the desired transformation was realized through a carefully chosen imperfect polarizing beamsplitter. We tilted the beamsplitter to fine-tune the extinction ratio.

The singlet state projection uses an in-fiber non-polarizing beamsplitter. The measured splitting ratio was 47:53. Unwanted polarization unitary transformations in the fiber and beamsplitter are compensated using a tilted quartz plate, a HWP, and a QWP placed in the arm without polarization-dependent loss, just before the entrance of the fiber to the non-polarizing beamsplitter. The output fibers of the beamsplitter traveled to single-photon detectors (avalanche photodiodes). A coincidence (i.e. simultaneous detections) signifies a successful projection on an $\ket{\text{MP}_i}$ state.

This implementation includes a $3/4$ loss due to the sequential implementation of Massar and Popescu state projections, as well as a $1-\eta(\mathcal{C}=0.25) = 49\%$ loss per projection. This results in an overall $\sim 12.7\%$ efficiency. 

Despite these efficiency limitations, the primary objective of this work is to demonstrate the feasibility, performance, and scaling properties of quantum state estimation with the Massar and Popescu entangling measurement on two particles. The insights gained will inform the viability and the level of performance that can be expected of this collective strategy when applied to useful systems in quantum science and technology involving multiple particles. 

\subsection{Optimal separable measurement strategy}
The optimal solution for the GenMP game when restricted to LOCCs can be achieved by measuring one photon in the $\{\ket{H}, \ket{V}\}$ basis and the other in the $\{\ket{D}, \ket{A}\}$ basis \cite{Massar2000}. The guessed state $\ket{\text{L}_i}$ on the unit sphere is taken to be at the bisector of the outcome states of the two photons:
\begin{equation}
\begin{aligned}
    &\ketbra{D}{D}\otimes \ketbra{H} \rightarrow \ket{\text{L}_1} = \cos \frac{\pi}{8} \ket{H} + \sin\frac{\pi}{8}\ket{V}, \\ 
    &\ketbra{D}{D}\otimes\ketbra{V} \rightarrow \ket{\text{L}_2} = \cos \frac{3\pi}{8} \ket{H} + \sin\frac{3\pi}{8}\ket{V},  \\
    &\ketbra{A}{A}\otimes\ketbra{H}\rightarrow \ket{\text{L}_3} = \cos \frac{\pi}{8} \ket{H} - \sin\frac{\pi}{8}\ket{V},  \\
    &\ketbra{A}{A}\otimes\ketbra{V} \rightarrow \ket{\text{L}_4} = \cos \frac{3\pi}{8} \ket{H} - \sin\frac{3\pi}{8}\ket{V}. 
    \label{eq:LOCC}
\end{aligned}
\end{equation}
 Experimentally, the LOCC optimal strategy was implemented by introducing a long delay between the photons in the two arms in Fig. \ref{fig:setup} (to avoid the HOM interference underlying the singlet projection) and measuring coincidences with polarizers placed in front of single-photon detecting avalanche photodiodes.

\subsection{Experimental test of the Massar and Popescu game} 
To test the implementation of the GenMP and TetraMP games, we considered a set $S$ of 40 states. These states were generated randomly from a uniform distribution $P(\vec{n}) d\vec{n}$ over the Poincare sphere by applying Haar random unitaries to the state $\ket{H}$. The states are of the form:
\begin{equation}
    \ket{\vec{n}} = \cos \theta/2 \ket{H} + e^{i\phi} \sin{\theta/2}\ket{V},
\end{equation}
where the polar angle $\theta \in [0, \pi]$ and the azimuthal angle $\phi \in [0, 2\pi)$ parameterize the state. The expected average fidelity for the set $S$ when using collective measurements is $\bar{\mathcal{F}}^{\text{Th, Coll}}_S = 75.05\%$, while the expected fidelity for the LOCC measurement is $\bar{\mathcal{F}}^{\text{Th, LOCC}}_S = 73.68\%$. We measured the average fidelity for each state based on $\sim 3500$ detected photon pairs, that is, $\sim 3500$ trials of $N=2$. The number of trials per state varied slightly in the experiment, although the same amount of time was spent on each state. These small variations are taken into account in the theoretical average fidelities $\bar{\mathcal{F}}^{\text{Th, Coll}}_S$ and $\bar{\mathcal{F}}^{\text{Th, LOCC}}_S$. To obtain the experimental average fidelity over all states, corresponding to Eq. \ref{avgFidExpression}, we calculate the weighted average:
\begin{equation}
\bar{\mathcal{F}}^{\text{Exp}}_S =\sum_{\ket{\vec{n}}\in S} P(\vec{n})\sum_{g=1}^4 P(g|\vec{n})\mathcal{F}(\ket{\vec{n}}, \ket{\vec{n}_g}).
\end{equation}
Although $P(\vec{n}) \approx 1/40$, we calculate $P(\vec{n})$ as the total coincidences measured for $\ket{\vec{n}}\otimes\ket{\vec{n}}$, divided by the total coincidences measured over all input states. This accounts for small variations in the number of trials per state. Now a frequency, $P(g|\vec{n})$ is calculated as the ratio of the coincidence counts for the projector corresponding to $\ket{\vec{n}_g}$ and the total coincidences for $\ket{\vec{n}}$. The guess $\ket{\vec{n}_g}$ is $\ket{\uparrow_{n_i}}$ (Eq. \ref{eq:tetras}) when using the collective strategy, and $\ket{\text{L}_i}$ (Eq. \ref{eq:LOCC}) when using the LOCC approach.

\section{Results}
\subsection{Experimental General Massar and Popescu game}

\begin{figure}[H]
    \centering
    \includegraphics[width=\linewidth]{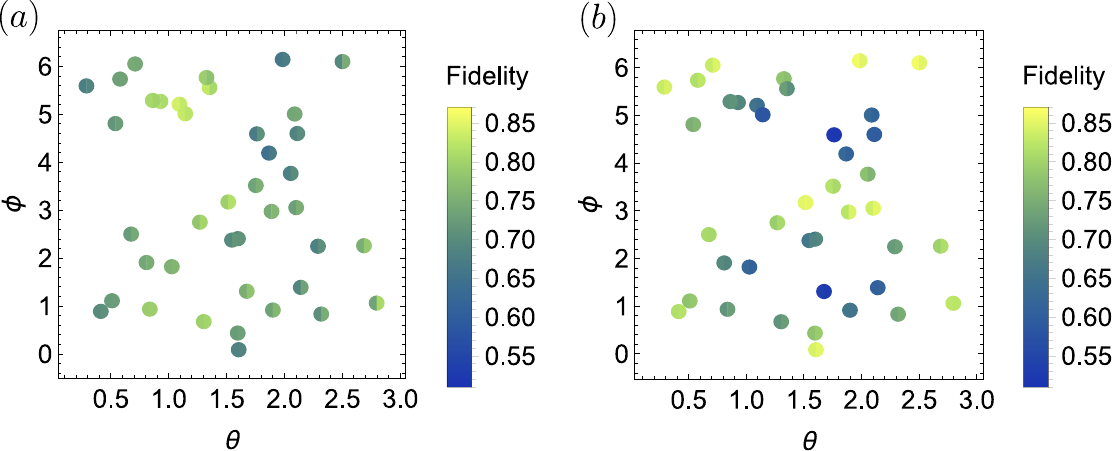}
    \caption{Scatter plots for the fidelity of the Massar and Popescu two-copy game: (a) using the collective approach; (b) using the LOCC approach; over the $\{\theta, \phi\}$ state-space. The left-side of each circle represents the experimentally measured fidelity, while the right-side represents the theoretically expected fidelity.}
    \label{fig:fidsMP}
\end{figure}

 In Fig. \ref{fig:fidsMP}, we compare the experimentally measured fidelity with the theoretically expected one for each state using the collective measurement in (a) and the LOCC approach in (b). The left side of each circle represents the experimentally measured fidelity, while the right side represents the theoretically expected fidelity. The experimental and theoretical values for the fidelity are generally in good agreement. The experimental average fidelities observed for the collective and LOCC schemes shown in Fig. \ref{fig:fidsMP} were $\bar{\mathcal{F}}^{\text{Exp, Coll}}_S = (73.5\pm0.2)\%$ and $\bar{\mathcal{F}}^{\text{Exp, LOCC}}_S = (73.4\pm0.2)\%$, respectively. The average fidelity using the collective strategy was found to be at least as high as that of the LOCC approach. 

The deviations from the expected fidelities can be in part attributed to imperfect HOM interference. When the state prior to the 47:53 beamsplitter was prepared as $\ket{V}\otimes \ket{V}$, the HOM visibility was $(98.7\pm0.4)\%$, while it was $(93\pm3)\%$ when the state entering the beamsplitter was $\ket{A}\otimes \ket{A}$. This suggests a slightly imperfect compensation of the polarization changes in single-mode fibers, which were used to ensure spatial-mode overlap for HOM interference. We can consider the effect of this additional polarization unitary $\tilde{U}$ that satisfies $\abs*{\expval{\tilde{U}}{V}}^2 = 98.7\%$ and $\abs*{\expval{\tilde{U}}{A}}^2 = 93\%$. Numerical optimization can be used to find the $\tilde{U}$ that minimizes the mean-squared error between the experimentally measured and theoretically expected fidelities for the states in $S$ using the collective strategy, subject to the constraints $\abs*{\expval{\tilde{U}}{V}}^2 = 98.7\%$ and $\abs*{\expval{\tilde{U}}{A}}^2 = 93\%$. The  theoretically  expected  average  fidelity  in  the  presence  of $\tilde{U}$ which  best  matches  the  experimentally  measured  values  is  about  $74.4\%$, compared  to  $75.0\%$  expected  in  its  absence. The gap suggests that  eliminating  these  systematic  errors  would  at  minimum increase  the  average  fidelity  by 0.5 \%. Using this simple model, the average fidelity $\bar{\mathcal{F}}^{\text{Exp, Coll}}_S$ would at least increase to $\bar{\mathcal{F}}^{\text{Exp, Coll}}_{S, \text{ Corrected}} = (74.0\pm0.2)\%$ in the absence of $\tilde{U}$. As the LOCC approach does not rely on HOM interference, $\bar{\mathcal{F}}^{\text{Exp, LOCC}}_S = (73.4\pm0.2)\%$ would remain unchanged. We see that when accounting for systematic errors, the collective strategy outperforms the LOCC approach. 

 \subsection{Experimental Tetrahedron Massar and Popescu game} 
\begin{figure}[H]
    \centering
    \includegraphics[scale=0.9]{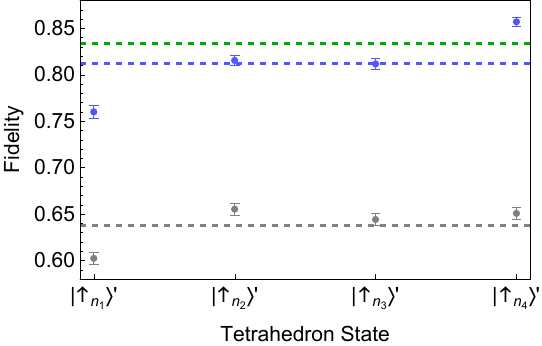}
    \caption{Fidelity of the Massar and Popescu two-copy game for the rotated tetrahedron states $\ket{\uparrow_{n_i}}'=\hat{R}\ket{\uparrow_{n_i}}$, where $\hat{R}$ is a fixed rotation in the Bloch sphere. These states are equivalent to Eq. \ref{eq:tetras} for the optimal collective measurement (see Supplemental document for details). The green dashed-line indicates the expected fidelity of $\bar{\mathcal{F}}^{\text{Coll}}_\text{Tetra} = 5/6 \approx 0.8333$. The experimental collective measurement results are presented in blue, while the experimental measurement results with suppressed entanglement are in gray. The dashed lines represent the average for each data set. The error bars are obtained by assuming Poisson statistics and using Gaussian error propagation.}
    \label{fig:fidsMPTetra}
\end{figure}

 We also experimentally examined the TetraMP game, with the measured fidelities for the rotated tetrahedron states presented in Fig. \ref{fig:fidsMPTetra}. The deviations in tetrahedron state fidelity from the expected value of $83.33\%$ can be in part attributed to the unitary $\tilde{U}$, which slightly distorts the measured frequencies by affecting the HOM interference statistics. These frequency deviations depend on the state prior to the beamsplitter rather than the prepared state $\ket{\vec{n}} \otimes \ket{\vec{n}} $. As a result, we are less likely to correctly guess the input $\ket{\uparrow_{n_1}}'$ and more likely to do so for the input $\ket{\uparrow_{n_4}}'$. This trend is also visible in Fig. \ref{fig:fidsMP}(a), where states with a significant overlap with $\ket{\uparrow_{n_1}}'$ (in the plot's bottom right) exhibit lower fidelities than expected, while states with substantial overlap with $\ket{\uparrow_{n_4}}'$ (with $\theta \approx 1.2$ and $\phi \approx 5.2$) show higher than expected fidelity.

 The average fidelity for the TetraMP game was found to be $\bar{\mathcal{F}}^{\text{Exp, Coll}}_\text{Tetra} = (81.2 \pm 0.3)\%$. This was compared with a version of the measurement, performed with a long delay between the photons, with no other changes to the setup. This corresponds to a complete suppression of the entangling component of the collective measurement. The average fidelity in this case was measured to be $\bar{\mathcal{F}}^{\text{Exp, Supp-Ent}}_\text{Tetra} = (63.8 \pm 0.3)\%$. This gap indicates the importance of the entangling measurement in approaching the optimal bound, improving the fidelity by $\sim 17\%$.

\subsection{Quantum state tomography using collective measurements} 
Projections onto the non-maximally entangled states $\ket{\text{MP}_i}$ can also be used for quantum state tomography. A large ensemble of size $N_{\text{ens}}$ (potentially of thousands of photons) can be divided into pairs on which we perform collective measurements to observe the tomographic efficiency. The estimate of the state is found using a maximum-likelihood reconstruction algorithm \cite{Shang}. Constraints regarding purity and the tensor-product form $\ket{\vec{n}}\otimes\ket{\vec{n}}$ of the input state are also imposed. 

\begin{figure}[ht]
\includegraphics[width=\linewidth]{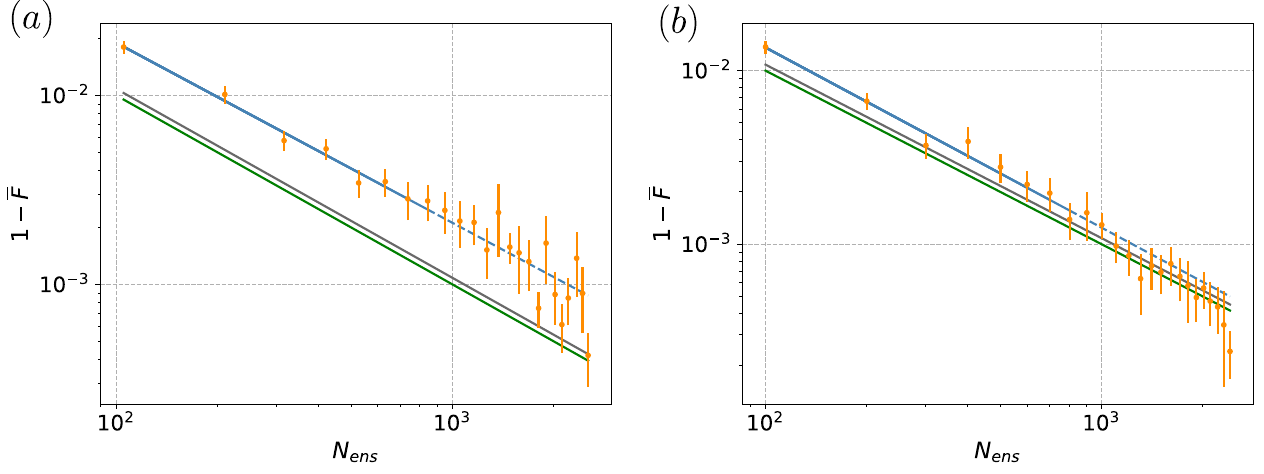}
    \caption{Tomographic infidelity with the number of samples $N_\text{ens}$ for the randomly-chosen states (a) $\ket{\vec{n}} = 0.883 \ket{H} + (-0.216 + 0.416 i) \ket{V}$ and (b) $\ket{\vec{n}} = 0.308 \ket{H} + (-0.916 + 0.256 i) \ket{V}$ shown on a log-log plot. The average values over multiple experimental trials for the collective measurement tomography are shown in orange, with the blue line showing the power-law fits to the first eight data points: (a) $1-\bar{F}=(1.5\pm0.4)N_\text{ens}^{-0.95\pm 0.04}$; (b) $1-\bar{F}=(1.6\pm0.4)N_\text{ens}^{-1.04\pm 0.05}$. The plotted error bars are calculated as the standard error of the mean over multiple trials. The green line indicates the Gill-Massar bound $1-\bar{F}=1/N_\text{ens}$, while the grey line indicates the theoretical average infidelity of Pauli tomography for pure states, given by $1-\bar{F}=(13/12)N^{-1}_\text{ens}$.}
    \label{fig:QST}
\end{figure}

To analyze the performance of the state tomography, the tomographic infidelity $1-\bar{F}$ is plotted as a function of the number of measured samples $N_\text{ens}$. The exact state that we produce will differ from the nominal state due to experimental imperfections. Thus, we compare our state estimate to the pure state estimate with the highest number of samples gathered ($N_\text{ens}\sim 15000$) instead of the nominal state.  Fig. \ref{fig:QST} shows the typical scaling of the infidelity observed for two randomly chosen states. Each data point is an average value over multiple reconstructions that used the same number of samples $N_\text{ens}$. To mitigate bias toward the reference state at higher $N_\text{ens}$ values, we applied a power-law fit to the initial eight data points (further details are provided in the Supplemental document).  We observed that the infidelity for each state approximately scales as $ O(1/N_\text{ens})$, the optimal infidelity scaling according to the Gill-Massar bound. The six random states examined had a combined average infidelity scaling of $1-\bar{F} = (1.7\pm0.2) N^{-(1.01\pm0.02)}_\text{ens}$, where the fit uncertainties for the states have been propagated. Infidelity scaling results for all tested states can be found in the Supplemental document. The coefficient could likely be improved with further attempts to minimize experimental imperfections and systematic errors. The optimal measurement (with $1-\bar{F} = 1/N_\text{ens}$) requires an $N_\text{ens}$-copy collective measurement of all the samples, but it is encouraging to see that the two-copy collective measurement tomography does achieve a near-optimal scaling.

It is worth noting that for exactly pure one qubit states uniformly distributed on the Bloch sphere, separable six-outcome Pauli tomography would theoretically allow an asymptotic average infidelity scaling of $(13/12) N^{-1}_\text{ens}$ \cite{Bagan2002,deBurgh}, shown in grey in Fig. \ref{fig:QST}. The use of either two-stage adaptive tomography \cite{Bagan2002} or random local von Neumann measurements \cite{bagan_comprehensive_2005} would improve the asymptotic average infidelity scaling to $1/N_\text{ens}$.  Therefore, a scaling advantage is not expected in this case. 

The scaling analysis could alternatively be performed with nearly-pure reference state estimates (purity $(99.91 \pm 0.02)\%$) obtained by averaging reconstructions with $N_\text{ens}\sim 2500$. The estimate is still constrained to being a pure state to comply with $\{\ket{\text{MP}_i}\}$, which are mainly useful for pure-state tomography. In this case, the six randomly tested states show a combined average infidelity scaling of $1-\bar{F} = (1.18 \pm 0.11) N^{-(0.93\pm0.02)}_\text{ens}$ (see Supplemental document for scaling plots). This suggests that the collective strategy is fairly robust in maintaining a near-optimal infidelity scaling, in contrast to a significant deterioration expected for Pauli tomography \cite{deBurgh, mahler_adaptive_2013}.

The goal in this part was to demonstrate the experimental feasibility of quantum state tomography using multi-particle collective measurements, and to show that it is comparable to the Gill-Massar bound in terms of average infidelity scaling.  Additionally, we observe this $\sim O(1/N_\text{ens})$ scaling with four outcomes as opposed to the six or more needed for the relevant Pauli, adaptive, and random tomography schemes \cite{Bagan2002, bagan_comprehensive_2005, deBurgh, huszar_adaptive_2012, kravtsov_experimental_2013 ,mahler_adaptive_2013}. 

\section{Conclusion}
We performed a proof-of-concept demonstration of Massar and Popescu's state estimation game using collective measurements on two particles. Our results suggest that the collective strategy can be in practice at least as effective as the best LOCC approach, and outperforms the local approach if systematic errors can be further reduced. They also emphasize the crucial role of entanglement in performing multi-particle collective measurements for tomography, with our entangling measurements greatly surpassing those where entanglement was suppressed. We also applied collective measurements to quantum state tomography, observing a near-optimal scaling of the infidelity with the number of measured samples. Our work opens the avenue to further implementations of collective measurements in the study of optimal strategies for parameter and state estimation tasks. 

Collective measurements could be particularly beneficial in systems where state-preparation is an expensive resource. Notably, optimal collective measurements for state estimation have been theoretically formulated when using: more than two particles \cite{derka_universal_1998, latorre_minimal_1998}, mixed qubit states \cite{vidal1999}, as well as qudit states \cite{brus_optimal_1999, hayashi_reexamination_2005}. The experimental implementation of optimal collective measurements on an arbitrary number of photons remains an open question. Practical challenges such as independent polarization control of multiple photons and reduced interference visibility in multi-port beamsplitters would need to be carefully considered.

\begin{backmatter}
\bmsection{Funding}
Canada First Research Excellence Fund; Natural Sciences and Engineering Research Council of Canada; Canada Excellence Research Chairs, Government of Canada.

\bmsection{Acknowledgments}
This work was supported by the Canada Research Chairs Program, the Natural Sciences and Engineering Research Council, and the Canada First Research Excellence Fund (Transformative Quantum Technologies).

\bmsection{Disclosures}
The authors declare no conflicts of interest.

\bmsection{Data Availability} Data underlying the results presented in this paper are available in Ref. \cite{mansouri2025demonstration}.

\bmsection{Supplemental document} See Supplement 1 for supporting content.

\end{backmatter}


\bibliography{allRefs}

\end{document}